\documentclass{article}

\usepackage{arxiv}

\usepackage[utf8]{inputenc} % allow utf-8 input
\usepackage[T1]{fontenc}    % use 8-bit T1 fonts
\usepackage{hyperref}       % hyperlinks
\usepackage{url}            % simple URL typesetting
\usepackage{booktabs}       % professional-quality tables
\usepackage{amsfonts}       % blackboard math symbols
\usepackage{nicefrac}       % compact symbols for 1/2, etc.
\usepackage{microtype}      % microtypography
\usepackage{lipsum}
\usepackage{graphicx}
\usepackage{natbib}

\hypersetup{%
  colorlinks=true,% hyperlinks will be coloured
  linkcolor=blue,% hyperlink text will be blue
  citecolor=blue,
  urlcolor=blue
}

\title{Submillimeter Galaxy studies in the next decade: EAO Submillimetre Futures White Paper Series, 2019}

\author{
  Ran Wang\thanks{\texttt{rwangkiaa@pku.edu.cn}}\\ 
  Kavli Institute for Astronomy and Astrophysics,
  No. 5 Yiheyuan Road, Haidian districut, 
  Beijing, 100871, China\\
   \And
  Wei-Hao Wang \\
  Academia Sinica Institute of Astronomy and Astrophysics (ASIAA), 
  No. 1, Section 4, Roosevelt Rd., Taipei 10617, Taiwan\\
  \AND
  David L. Clements\\
  Astrophysics Group, Blackett Laboratory, Imperial College London, London SW7 2AZ, UK \\
  \And
  Haojing Yan \\
  Department of Physics and Astronomy University of Missouri Columbia, 
  MO 65211, USA \\
  \And
  Yiping Ao \\
  Purple Mountain Observatory, Chinese Academy of Sciences, 
  Nanjing 210034, China \\
}

\begin{document}
\maketitle

\begin{abstract}
Over the last two decades, the Submillimetre Common-User Bolometer
Array (SCUBA) and SCUBA-2 on the James Clerk Maxwell Telescope (JCMT) achieved gread success in 
discovering the population of dusty starburst galaxies in the early universe. The SCUBA-2 surveys 
at 450 $\mu$m and 850 $\mu$m set important constraints on the obscured star formation over cosmic time, 
and in combination of deep optical and near-IR data, allows the study of protoclusters and structure 
formation. However, the current submillimeter (submm) surveys by JCMT are still limited by area of sky coverage 
(confusion limit mapping of only a few $\rm deg^2$), which 
prevent a systematic study of large samples of the obscured galaxy population. In this white paper, 
we review the studies of the submm galaxies with current submillimeter/millimeter (submm/mm) observations, 
and discuss the important science with the new submm instruments in the next decade. In particular,  
with a 10 times faster mapping speed of the new camera, we will expect deep 850 $\mu$m surveys 
over 10 to 100 times larger sky area to i) largely increase the sample size of submm 
detections toward the highest redshift, ii) improve our knowledge of galaxy and structure formation in the early universe.

\end{abstract}

% keywords

\section{Introduction}
\label{sec:intro}

The thermal dust continuum emission at far-infrared wavelengths is an important trace of the dust and 
gas contents and star forming activities in galaxies \cite{c06,s14,g17}. At high redshift, the UV and optical emission 
from the stellar component is dimming dramatically. However, the hump of the thermal dust emission is 
shifted to the submm bands, and due to the negative $k$-correction, the observing flux densities 
at submillieter and millimeter wavelengths do not drop with redshift. Thus, the submm/mm windows open a  
unique opportunity to probe active star formation and galaxy evolution toward the earliest epoch. 
Over the last two decades, the submm facilities, such as the Submillimetre Common-User Bolometer
Array (SCUBA) and SCUBA-2 on the JCMT and the SPIRE on the \emph{Herschel} Space telescope etc., achieved
gread success in discovering the population of dusty starburst galaxies in the early universe \cite{h98,g17,v14}.

Dusty starbursts have been found into the epoch of reionization (EoR; $z>6.3$).
Some of them are found as quasar hosts and/or companions \cite{b03,d17}, and yet some are found in ''blind'' sub-mm/mm
surveys \cite{r13,s17}. In terms of IR
luminosities, they are all ULIRGs ($L_{IR}>10^{12}L_\odot$) and HyLIRGs
($L_{IR}>10^{13}L_\odot$), which translate to dust-embedded star formation rates
(SFRs) of $>100-1000 M_\odot/yr$. The very existence of such high-z U/HyLIRGs
has important implications. The burst of star formation traces the most active stage of 
galaxy evolution. The submm sources detected in the core of overdensity regions also probe 
the early evolution of galaxy clusters. In addition, The prevalence of dust at $z\approx 6$-7 
means that there must be very active star formations at even earlier epochs ($z\sim 10$ and beyond). 

Bright submm/mm emission was also detected in the host galaxies of optically 
luminous quasars at z$\sim$2 to 7 \cite{o01,o03,p03,w07,l13}. The single dish surveys at sub-mJy sensitivity 
reveal that about 30\% of them are hosted in dusty starburst systems with FIR
luminosities comparable to that of the ULIRGs and HyLIRGs, suggesting massive 
star formation co-eval with rapid supermassive black hole (SMBH) accretion. 
These quasar-starburst systems became important targets for further interferometer 
observations to search and resolve the gas and dust content. The follow-up 
molecular CO and [C II] observations probe the distribution of dust, gas, and star 
forming activity, as well as the host dynamics \cite{m05,c07,w13,d17,d18,v16}, and thus, set key constraints 
on the early growth of the SMBH-galaxy systems \cite{v10,va13}.     

The obscured star formation discovered in submm surveys provides essential complement to probe the cosmic star formation history (SFH). 
Traditionally, tracing the evolution with cosmic time of the galaxy luminosity
density from the far-UV (FUV) to the far-infrared (FIR) offers the prospect of
an empirical determination of the global star formation history (SFH) and heavy
element production of the Universe \cite{m14,r16,l17}. However, lacking
precise astrometry at FIR wavelengths at high
redshifts often prevents us from connecting FUV to FIR data and providing the
required complete understanding of cosmic SFH.
To study the formation and evolution of galaxies, it is crucial to determine
the redshift distribution of sources. Large samples of high redshift galaxies
have been imaged with space facilities like \emph{Hubble}, \emph{Spitzer}, \emph{Herschel} and
ground-based telescopes at multiple wavelengths and their redshifts can be
determined by photometric observations at multiple bands. Alternatively, a
small fraction of sources have been confirmed with spectroscopic measurements
and narrow-band imaging. Due to large extinction, observations at optical
and/or near-infrared are difficult and only bright sources can be detected at
high redshifts. For some sources with large amount of dust, they are very
likely to be invisible with current optical facilities in a limit observing
time. 

The mapping capability of the single dish telescopes at submm wavelengths are also 
becoming important in tracing overdensity and structure formation in the early universe.
Clusters of galaxies are the most massive bound structures in the local universe. They are largely dominated 
by `red and dead' elliptical galaxies, and the oldest and most massive elliptical galaxies lie at their 
cores (eg. \cite{s12}). Studies of the stellar populations of these massive galaxies reveal them to be old, 
with inferred formation redshifts $z>3$ and with the bulk of their stellar populations forming over a short 
time scale. This would imply that the progenitors of the massive elliptical galaxies in the cores of clusters 
must have formed in major starbursts.

Finding galaxy clusters in formation at these high redshifts is a difficult task. The standard methods of cluster 
detection include X-ray and Sunyaev Zel'dovich (SZ) observations of the hot intracluster medium, and the search for red sequence galaxies 
in the optical and near-IR. All of these methods fail for forming galaxy clusters; the first two since young systems 
are yet to Virialize and thus lack a significant intracluster medium that can be detected by X-rays or SZ, and the last 
because the galaxies making up the cluster are still star forming and thus do not lie on the red sequence. The high 
star formation rates (SFRs) of forming giant elliptical galaxies, however, in principle allow us to search for these 
objects in the far-IR, since, like starbursts in the local universe, they should be luminous at these wavelengths. 
Recent results using {\em Herschel} and {\em Planck} data in the far-IR and submm have begun to find candidate protoclusters in this way.

\section{Current Status}
\label{sec:current}

Submm surveys of galaxies were carried out with JCMT in the last two decade (Table 1), e.g., the SCUBA program which discovers the 
first submm galaxy sample at high redshift, the SCUBA-2 Cosmology Legacy Survey (S2CLS, \cite{g17}), 
The Hawaii SCUBA-2 Lensing Cluster Survey \cite{h16}, the Submillimeter Perspective on the GOODS Fields (SUPER GOODS, \cite{c17}),
the SCUBA-2 Large eXtragalactic Survey (S2LXS\footnote{https://www.eaobservatory.org/jcmt/science/large-programs/s2lxs/}), S2COSMOS/eS2COSMOS\footnote{https://www.eaobservatory.org/jcmt/science/large-programs/s2-cosmos/}, 
and SCUBA-2 Ultra Deep Imaging EAO Survey (STUDIES, \cite{w17}). The SCUBA-2 surveys image the submm 
sky to a 1$\sigma$ noise level of $0.9\sim2$ mJy, discovering the extreme starburst systems with FIR luminosities 
on orders of $\rm 10^{12}$ to $\rm 10^{13}\,L_{\odot}$. These observations, together with the data from \emph{Spitzer} and 
Herschel at shorter wavelengths, measure the submm source number densities and update the star formation rate 
density \cite{h98,d05,g13,r16,g17,w17}, revealing that the submm galaxies contribute significantly 
to the cosmic star formation history over a wide range of redshift \cite{m14,r16}. 

To study the SFH in the early Universe, observations by space
telescopes at far-infrared and ground facilities at submm become an important
tool to detect highly dust-obscured galaxies. However, due to the poor
resolution and confusion level of the space telescopes, the observations will only select bright sources and therefore largely
underestimate the SFR at high redshifts. Ground-based facilities like JCMT/SCUBA2
can provide better sensitivity and resolution in comparison with space
facilities like \emph{Herschel}. However, due to a relatively low mapping speed the
statistic results still suffer large uncertainties by the cosmic variance from
small surveyed fields. Large area of deep surveys are highly required in
future.

In addition, to fully understand the early cosmological star formation history, we will need
large samples of high-z dusty starbursts selected in a systematic manner. The
most promising method to construct such samples is through the use of
FIR/sub-mm colors in blind, un-biased surveys. The typical cold-dust emission
(as being heated by star formation) in galaxies has its peak at around
rest-frame 100~$\mu$m, and the color selection is to utilize this
characteristics. The so-called ``500 $\mu$m riser'' technique is an
implementation tailored for the BLAST and the \emph{Herschel}/SPIRE bands \cite{p10,r12}: a dusty galaxy at $z>4$ should have
red colors in 250, 350 and 500~$\mu$m bands because its SED is ``rising''
through these three bands as the peak is redshifted to $\sim 500$~$\mu$m and
redder wavelengths. An extension of this technique to higher redshifts is the
``850/870~$\mu$m riser'' method, where a redder band at 850 or 870~$\mu$m
is added to the selection \cite{r17}.

\begin{figure}
\centering
\includegraphics[width=0.7\textwidth]{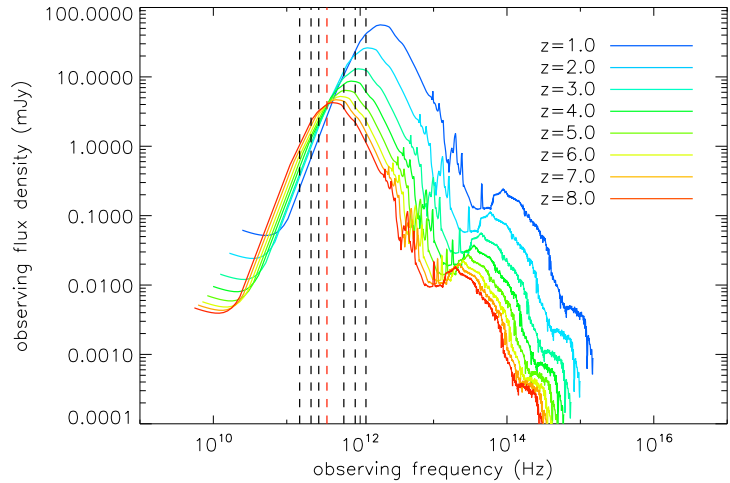}
\caption{The spectral energy distribution (SED) of a dusty star forming galaxy at redshift from z=1 to 8. We use 
the SED of M82 from the SWIRE library as a template \cite{p07}. 
The SEDs are normalized at an 850 $\mu$m observing flux density of 4mJy, which are the typical 4$\sigma$ detection 
limit for JCMT/SCUBA-2 surveys. The vertical lines shows the observing 
windows at 250 $\mu$m, 350 $\mu$m, 500 $\mu$m, 850 $\mu$m, 1.1 mm, 1.4 mm, 2 mm. The 850 $\mu$m window is 
marked as a red line, which samples the peak of the dust continuum bump at the highest redshift.}
\end{figure}

A growing number of protoclusters and protocluster candidates are being found 
using far-IR and submm techniques. Casey \cite{c16} summarises results on five specific 
protoclusters at $2<z<3.1$ with spectroscopic confirmation, as well as a further three candidates 
at higher redshifts. Meanwhile, there is an increasing number of candidate protoclusters with 
estimated redshifts around 2 to 3 emerging from work on the {\em Herschel} and {\em Planck} 
surveys, including at least 27 candidates emerging from investigating {\em Planck} sources lying 
within the large area {\em Herschel} surveys \cite{g18}, \cite{c14}, and 228 resulting from 
colour selection from the {\em Planck} \cite{p16} all sky survey. The nature of these sources is 
currently unclear since nearly all the sources lack the spectroscopic followup necessary to confirm, 
or otherwise, their protocluster nature. One of the {\em Planck} candidates has been found to be 
a superposition of two galaxy overdensities at different redshifts (1.7 and 2.0, \cite{f16}) while 
photometric redshifts suggest that several others are genuine protoclusters (\cite{c14}, \cite{c16}). 
Over the next ten years the spectroscopic study of these candidates should allow considerable 
progress in confirming their nature and studying the detailed physics of the processes driving 
the star formation in their member galaxies.

\begin{figure}
\includegraphics[width=0.9\textwidth]{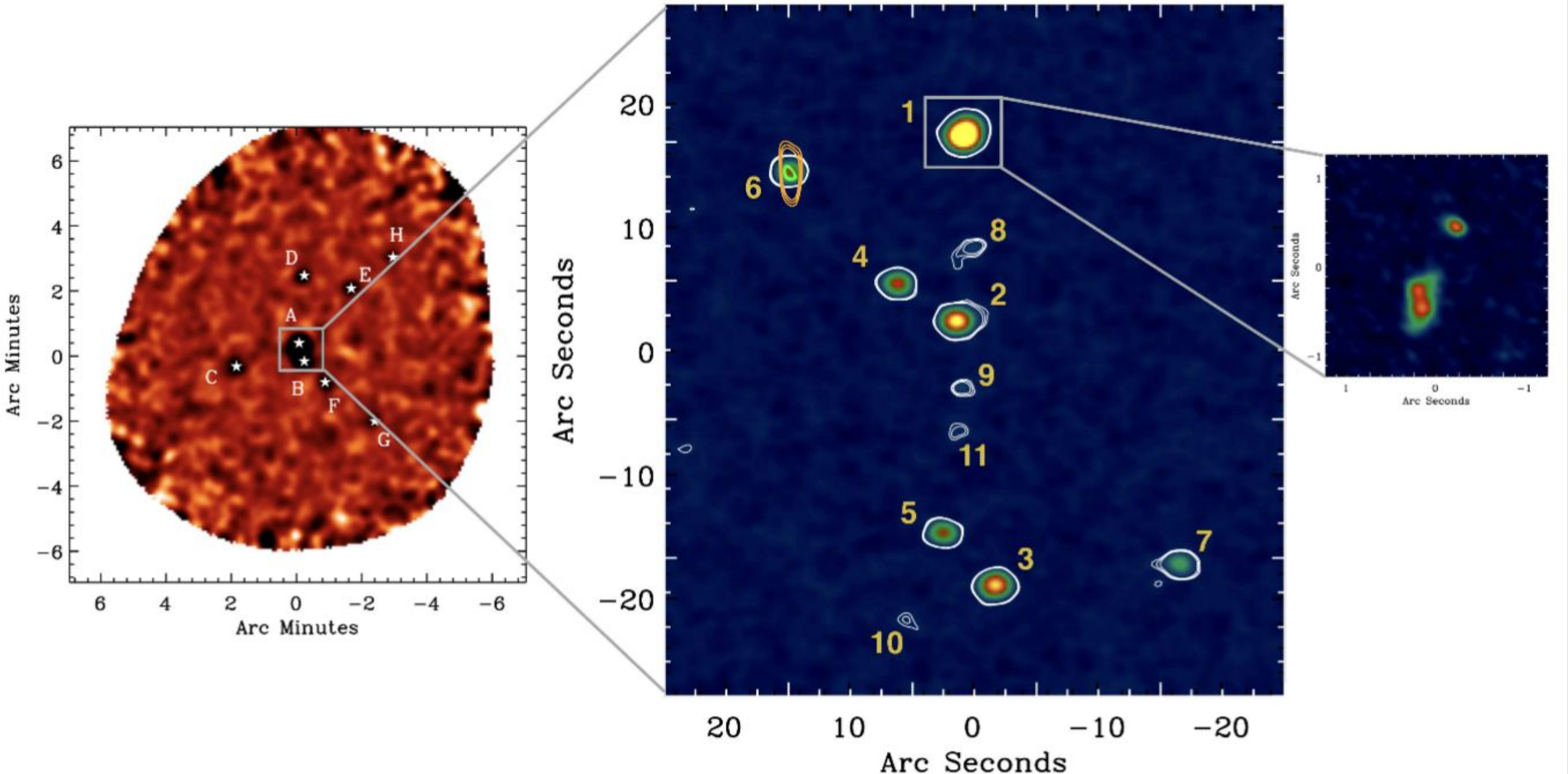}
\caption{From \cite{o18}: A candidate protocluster core at a redshift of 4.001 discovered through colour selection from {\em Herschel} and followup submm imaging. On the left is a LABOCA image of a very red {\em Herschel} source selection from its 250 to 350 to 500 $\mu$m colour. The {\em Herschel} source breaks up into several 870$\mu$m sources. In the centre is an ALMA image of the central region of this clump of sources, showing that they too break up into numerous submm sources. These sources have spectroscopic redshifts measured by ALMA. On the right is the brightest of these sources which, in higher reslution ALMA data, is found to be a pair of submm sources. This is consistent with the picture of \cite{c17} for protocluster core formation, but it is found at $z=4$ rather than the predicted $z\sim 6$ for this stage of cluster formation. For details see \cite{o18} and references therein.}
\end{figure}

Theoretically, protoclusters at $z\sim$2 are expected to be very large structures. Theoretical models 
show that the eventual members of a Coma-sized cluster at zero redshift will be spread over scales 
of 10s of Mpc \cite{m15}. Observations of several confirmed protoclusters (\cite{c15}; \cite{y14}; \cite{d14}) 
seem to confirm this result, with structures seen on scales of 3 to 15 Mpc. However, there is a mismatch between 
the detailed predictions of SFRs and other properties for protocluster galaxies from theoretical models and 
what we appear to be seeing in the population revealed by {\em Herschel} and {\em Planck} (see eg. \cite{g18}). 
At higher redshifts, theoretical predictions suggest a different picture, with the cores of eventual protoclusters 
showing high rates of star formation on scales of a few 100kpc at $z\sim6$ \cite{c17}. The observational situation 
at higher redshift is somewhat confused, however. Starbursting cluster cores may have been seen at $z\sim4$, rather 
lower than the predicted redshift, in followup observations of very red sources from {\em Herschel} \cite{o18} and 
the South Pole Telescope \cite{m18}, but protocluster candidates at $z\sim6$ selection through Ly$\alpha$ emission by 
HyperSuprime Cam \cite{h19} show structures, including far-IR luminous sources, extended on scales of 10 Mpc instead.
It is thus clear that much remains to be learnt about the early stages of galaxy cluster formation.

\begin{figure}
\includegraphics[width=0.9\textwidth]{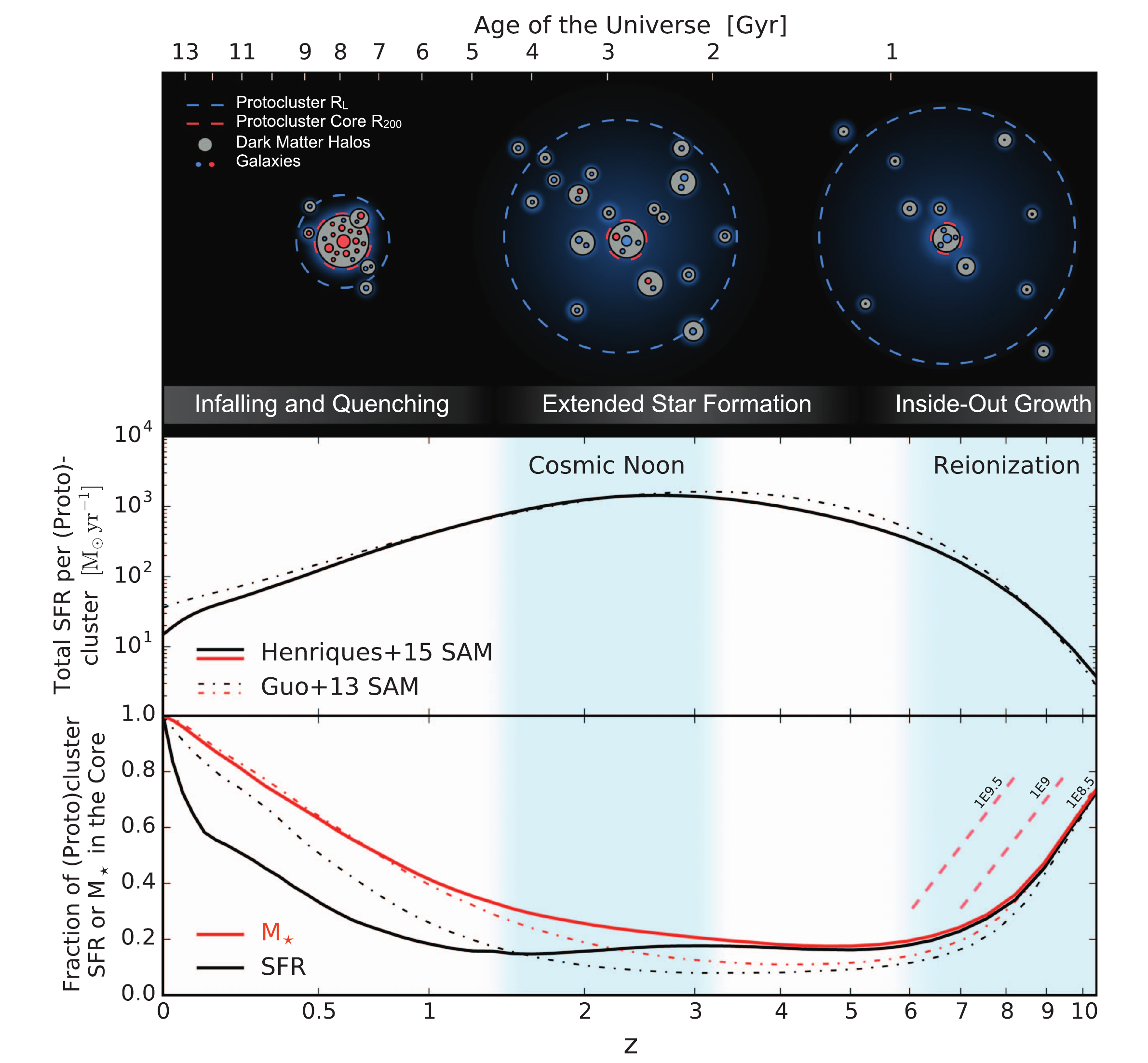}
\caption{From \cite{c17}: predictions for the evolution of protoclusters showing the core formation phase at high redshift and the growth phase at $z= 2-3$. For details see \cite{c17} and references therein.}
\end{figure}

In summary, the submm/mm bands are unique in tracing obscured star formation and galaxy evolution over the cosmic time.
The great success of the existing JCMT/SCUBA and SCUBA-2 survey proved that the 850 $\mu$m band on MaunaKea is the 
most efficient window for deep imaging to detect the submm population at high redshift. However, 
the current submm surveys by JCMT are limited by the small area of sky coverage (Table 1). 
e.g., The deep SCUBA-2 surveys (S2CLS, S2COSMOS) cover only $\leq 5$ deg$^{2}$ of sky area. For comparison, 
much large sky area are aleardy covered with deep optical, near-infrared, and radio 
observations (e.g., Stripe 82 of 300 deg$^2$). The lack of deep submm data in these region prevent a 
systematic study of large samples of the obscured galaxy population. Thus, the new JCMT 850 $\mu$m 
camera with a 10 times faster mapping speed becomes an urgent request for developing large submm surveys.   

\begin{table}[h!]
\caption{Submillimeter and millimeter surveys}
\centering
\begin{tabular}{lcccc}
\noalign{\smallskip} \hline \noalign{\smallskip}
Survey   & Sky Area     & Wavelength & 1$\sigma$ rms & Reference \\
         &  deg$^2$     & $\mu$m     & mJy beam$^{-1}$   &     \\
\noalign{\smallskip} \hline \noalign{\smallskip}
SCUBA/JCMT SHADES & 0.2 & 850 & 2 & \cite{c06}\\
SCUBA/JCMT HDF & 0.0025 & 850 & 0.45 & \cite{h98} \\
SUPER GOODS & 0.125 & 850/450 & 0.28/2.6 & \cite{cowie17} \\
Hawaii SCUBA-2 Lensing Cluster Survey & 0.137 & 450 & 4.4$^c$ & \cite{h16} \\
S2CLS & 5 & 850 & 1.2 & \cite{g17} \\
S2COSMOS/eS2COSMOS & 2 & 850 & 0.9 & \cite{s19} \\
STUDIES & 0.08 & 450 & 0.55 & \cite{w17} \\
S2LXS & 10 & 850 & 2 & a \\
HerMes &  380 & 250/350/500 & 5.2$\sim$12.8 & \cite{o12} \\
H\_ATLAS & 570 & 250/350/500 & 9 & \cite{e10} \\
HerS  & 79 & 250/350/500 & 13.0/12.9/14.8 &  \cite{v14} \\
BGS-Wide & 10 &  250/350/500 & 36/31/20 & \cite{c11} \\
BGS-Deep & 0.9 & 250/350/500 & 11/9/6 & \cite{c11} \\
LESS & 0.25 & 870 & 1.2 & \cite{w09} \\
AzTEC GOODS-S & 0.075 & 1100 & 0.48$\sim$0.73 & \cite{s10}  \\
AzTEC GOODS-N & 0.068 & 1100 & 0.96$\sim$1.16 & \cite{p08} \\
AzTEC SHADES & 0.7 & 1100 & 0.9 & \cite{a10} \\
AzTEC COSMOS & 0.72 & 1100 & 1.26 & \cite{a11} \\
SPT & 2500 & 1400/2000/3000 & 4/1.2/2 & \cite{m13,vi13} \\
IRAM/GISMO survey & 0.07 & 2000 & 0.23 & \cite{m19} \\
NIKA2 GOODS-N$^{d}$ & 0.04 & 1200/2000 & 0.2$\sim$0.6/0.1$\sim$0.3 & \cite{d16}\\
NIKA2 COSMOS$^{d}$ & 0.4 & 1200/2000 & 0.1/0.07 & \cite{d16}\\
\noalign{\smallskip} \hline\noalign{\smallskip}
\multicolumn{5}{c}{Planned surveys}\\
\noalign{\smallskip} \hline\noalign{\smallskip}
TolTEC Ultra-Deep Galaxy Survey & 1 & 1100/1400/2000 & 0.026/0.024/0.018 &b \\
TolTEC Large Scale Structure Survey & 100 & 1100/1400/2000 & 0.26/0.24/0.18 &b \\
\noalign{\smallskip} \hline
\end{tabular}\\
$^{a}$Geach et al. M17BL001; $^{b}$http://toltec.astro.umass.edu/science\_legacy\_surveys.php
$^{c}$Before lensing amplification correction.
$^{d}$See also Beelen et al. https://lpsc-indico.in2p3.fr/Indico/event/1765/session/9/contribution/52
\end{table}

\section{The Next Decade}
\label{sec:future}

We would like to propse wide field surveys at 850 $\mu$m using the new camera on the JCMT when the 
instrument is available in 2022. The 10 times faster mapping speed 
of the new 850 $\mu$m camera will allow submm mapping over hundred deg$^{2}$ of sky area at mJy sensitivity 
level in the next decade. The fields covered by the deep X-ray, optical, near-infrared, and radio observations will 
have the highest priority for such submm surveys. As we discussed above, the previous submm/mm surveys with sky 
coverages of 100 to 1000 deg$^2$, such as the HerS and HerMes Surveys at 250$\mu$m, 350$\mu$m, and 500$\mu$m, the SPT 
survey at 1.4mm, 2mm, and 3mm, are insufficient in both sensitivity and angular resolution, to fully recover the 
obscured star formation population toward the highest redshift and identify the counterpart of galaxy samples 
and protocluster members discovered in deep optical and near-IR observations. 
The new 850 $\mu$m continuum survey will largely increase the sample size of submm sources, 
and combined with data at other wavelengths, will significantly improve our knowledge of 
galaxy and structure evolution in the early universe.       

\subsection{Selection of Dusty Starbursts at Very High Redshifts}

High-z dusty starbursts are rare. Depending on the exact color criteria adopted,
the surface density of 500~$\mu$m risers in the HerMES and the H-ATLAS areas
is $\sim 10$~deg$^{-2}$ or less. While there is not yet sufficient statistics 
of 850/870~$\mu$m risers, there is evidence that they are even rarer (e.g.
Duivenvoorden et al. 2018). By providing an increase of $10\sim20\times$ in the
mapping speed than SCUBA-2, the planned JCMT new wide-field camera will be the
most powerful tool in our search for dusty starbursts at $z>7$ and beyond.
Furthermore, the increase in mapping speed will bring JCMT to the same league
as the new millimeter camera, TolTEC, on the LMT \cite{b18}.  Combining wide-area submm/mm data
from JCMT and LMT will greatly boost our capability to study the high-redshift 
dusty population in terms of both sample size and robust sample selection.

\subsection{Understanding the connection between star formation and SMBH accretion}

Large populations of quasars and active galactic nuclei (AGN) are discovered from the 
optical and near-IR surveys. Previous submm/mm observations discovered a small 
fraction of the quasar host galaxies that are still efficiently forming stars.   
However, the connection between star formation activity and AGN properties are still unclear. 
In particular, the limit sky coverage of previous submm/mm observations cannot provide enough sample 
in different bins of AGN activities for 
such studies at high redshift. The new 850 $\mu$m survey will increase the 
sample size of submm observed quasars by factors of $>$a few tens and significantly 
improve the statistics. 

The detection of dust continuum at 850 $\mu$m will intermediately constrain the dust mass  
and star formation rate in the quasar host galaxies. The sky coverage of large quasar 
samples will allow detections of the stacking signals with objects in bins of redshifts, SMBH masses, 
and quasar luminosities. This will address the evolutionary connection between star formation and 
SMBH accretion over a wide range of redshift. Based on the assumption of dust-to-molecular gas ratio, the gas 
mass could also be constrained. The gas fraction in quasar hosts is an important parameter to 
address the effect of AGN feedback (e.g., \citep{h08,s18}). i.e., a low gas fraction compared 
to the normal star forming galaxies may indicate that the molecular gas are 
removed and the star formation are suppressed due to the AGN power. In addition, 
More quasar-starburst systems will be discovered at high redshift. These objects will 
be the targets for further ALMA and JWST observations to image the dust, gas, and stellar 
components. These systems will be the key examples to probe the early growths of 
the SMBH and their host galaxies. 

\subsection{Cosmic star formation history based on the large, deep surveys}

Large samples of SMGs have been detected with the JCMT/SCUBA-2 (e.g., \citep{g17}), and follow-up observations with ALMA have mapped some 
SCUBA-2 sources.  Machine-learning algorithm can efficiently identify the likely
counterparts at optical/NIR for the SCUBA-2 sources by using ALMA observations
as a training sample \cite{a18}.  A new bolometer camera at JCMT with a
mapping speed of 10 times faster than the current SCUBA-2 can be used to finish
a much wider field with multiple optical/NIR archival data in an efficient
way.  Together with the machine-learning method, this can well constrain the
cosmic SFH and  significantly reduce the cosmic variance for the
measurements.

However, it is difficult to efficiently search for high redshift sources even
with ALMA.  Currently, lacking of identified high redshift sources will largely
underestimate their corresponding SFRs at z $>$5. ALMA observations show some
SCUBA-2 sources without any optical/NIR counterparts. The latter could be good
high-redshift candidates. The deep continuum observations with single dishes
can reveal a large sample of SMGs and the follow-up high resolution
observations with the interferometers like ALMA can locate their accurate
positions. Together with archival multi-wavelength data and possible follow-up
deep observations at optical/NIR with large optical/NIR telescopes or on-going
facility like TMT, one can constrain the cosmic SFH at high redshift.  The
machine-learning method is also helpful to identify the likely high-redshift
candidates without any optical/NIR counterparts. It may provide informative
clues or constraints on the cosmic SFHs at high redshift when ultra-deep
optical/NIR observations are currently not available yet.

To compare with the cosmology simulation, we always require a large survey area
to reduce the cosmic variance and to find some extreme sources or extreme
environments, which are important astrophysical laboratories. The number
density of massive galaxies at high redshifts can be use to test different
galaxy evolution models, which predict massive galaxies decline very rapidly at
$z$$>$4.

Other large area surveys with IRAM30m/NIKA2 and LMT/TolTEC will
provide deep images at mm wavelength, reaching a superior
sensitivity. Together with the data from the next-generation of the 850 $\mu$m camera at JCMT,
one can efficiently select high redshift candidates based on the color
criterion.  Ultra-deep radio surveys at LOFAR and SKA precursors can be
important for getting cross identifications and accurate positions due to their
superior sensitivities and large field of views in comparison to the submm
surveys.

\subsection{Protoclusters and the galaxy cluster formation}

The far-IR/submm based protocluster surveys discussed above were not designed for protocluster work. They have found protocluster candidates 
with total SFRs $>10,000 M_{\odot}$/yr, with individual member galaxies forming stars at rates of 100s to 1,000s of $M_{\odot}$/yr, 
but it is likely that they are seeing only the peak of the luminosity function both of protoclusters and of galaxies 
within protoclusters because of the relatively limited sensitivity of these surveys to sources at such high redshifts. 
Meanwhile, dependence on selection at short submm wavelengths, typically 350, 500 or 550 $\mu$m for the {\em Herschel} 
and {\em Planck} surveys, hampers studies of the highest redshift protoclusters at $z \geq 6$. At the same time, the 
existing samples have been selected using a range of rather heterogeneous methods, making statistical assessments of 
luminosity functions and evolution rather difficult.

The next decade will see a range of developments that will move these studies forward. Firstly, spectroscopic followup 
of the existing samples will significantly increase the number of confirmed protoclusters known at $z\geq 2$. This will 
be achieved using both mm/submm spectroscopy using ALMA and NOEMA, and optical/near-IR observations with instruments 
like KMOS and MUSE. Secondly, theoretical models for these objects, which require detailed n-body-hydro codes with high 
spatial resolution,  will improve our insight into this population. Thirdly, large area surveys at mm wavelengths using 
instruments such as NIKA2 and TOLTEC will provide new candidate protoclusters for detailed examination. It is here where 
JCMT will be able to contribute, since the addition of large area surveys at higher frequency submm wavelengths will greatly 
enhance our ability to select candidate protoclusters from $z\sim 2$ to the highest redshifts. The protocluster candidates 
we currently know have an area density of about 1 per 40 sq. degrees, so the necessary surveys will have to be large area and 
ideally reaching a sensitivity of $\sim$ 1mJy. Such surveys will be sensitive to not only the population we already know but 
also fainter protoclusters and protocluster galaxies, allowing us to examine the luminosity function of these objects. The 
proposed 10x enhanced mapping speed over SCUBA-2 will make a few hundred sq. deg. 850 $\mu$m survey to these sensitivities 
possible as part of a large area Legacy Survey, which will also be useful for many other studies. At the same time a larger 
field submm instrument will be needed to survey the $\sim$15 Mpc region around known protoclusters to search for starbursting 
galaxies in their infall region that, if the predictions of Muldrew et al. \cite{m15} are current, will subsequently fall 
into the clusters.  Other instrumentation developments, such as KIDS-based submm imaging spectrometers, able to measure 
redshifts for all submm sources in a field simultaneously, would be ideal to followup the protocluster candidates detected 
in these surveys, and to characterise the molecular and atomic gas properties of their member galaxies, allowing this population 
to be confirmed and analysed much more rapidly than is possible with current instruments.
The JCMT thus has a huge potential for studying rare submm emitters, such as protoclusters, using the proposed new instrumentation.

\subsection{Further request of the 450$\mu$m capability}

Surveys at 850 $\mu$m will play a leading role in discovering the dusty star 
forming sources. However, detections at a single submm band is insufficient to 
constrain the nature of the detections. Measurements of dust temperature and IR luminosity, as well as 
the photometric-redshift require observations at more submm/mm bands. For objects at z$\geq$3, the 450$\mu$m 
window samples the peak of the thermal dust emission (Figure 1), thus is critical in determine the dust SED. 
The \emph{Herschel} data at 350$\mu$m and 500$\mu$m are confusion limited and are 
only available for a small fraction of the sky (HerMes, H\_ATLAS, and HerS fields). 
The 450$\mu$m window at Maunakea and the 450$\mu$m capability of JCMT will remain unique.  JCMT has an angular resolution
at 450$\mu$m that is similar to that of the LMT in the millimeter.  Therefore, JCMT can detect 450 $\mu$m sources
that are much fainter than the \emph{Herschel}/SPIRE limits and the 850 $\mu$m limit of JCMT.
The imaging capability of JCMT at 450$\mu$m will become a very important complement for future 
multi-bands surveys. It will also provide a unique band for the color selections of the dusty 
objects toward the highest redshift, and for probing the normal galaxy population at the peak epoch ($z\sim2$) of
the cosmic star formation and AGN.

\subsection{Time and sensitivity requests}
With the new 850 $\mu$m wide field camera, we will be able to carry out surveys that cover very wide areas and still achieve very deep sensitivities. For example, we could conduct a survey over 300 deg$^{2}$ to $rms=2.0$~mJy in 2500 hours under the Band 2/3 weather condition with matched-filter applied (assuming that $\tau_{225GHz}=0.08$, and that the new camera is 10 times faster than that of the current SCUBA-2 in PONG3600 mode with 660 pointings). We could choose the fields that have already been covered by the previous Herschel surveys and/or will be covered by the coming LMT/TolTEC surveys. The much larger survey area compared to those of the existing SCUBA-2 surveys will allow us to select a very large sample (on the order of $\times 10^4$) of 850 $\mu$m risers that are candidates of very high redshift dusty starburst galaxies. The combination of 850 $\mu$m data with the Herschel, NIKA2, and TOLTEC data will also allow the selection of protocluster candidates. Moreover, the large sky area will cover a significant number of optically-selected quasars. According to the Sloan Digital Sky Survey fourteen data release of quasar catalog, about 5000 quasars at 2$<$z$<3$ (i.e., the peak of SMBH and galaxy evolution) are expected within 300 deg$^{2}$ \cite{p18}. While only a small fraction of them would be directly detected, the shear number of the undetected ones will provide enough stacking signal to obtain the average dust continuum emission and dust mass of the hosts in different SMBH mass and quasar luminosity bins. Meanwhile, deeper surveys could be carried out over tens deg$^2$ down to the 0.7$\sim$1 mJy level. As an example, a point source sensitivity of $rms=0.7$~mJy over 10 deg$^2$ will need 720 hours in Band 2/3 weather ($\tau_{225GHz}=0.08$). This is much deeper than the existing Herschel surveys, allowing us to detect dusty star forming galaxies or quasar hosts with FIR luminosities around $\rm 1\times10^{12}\,L_{\odot}$ at high redshifts. In addition, an area on the order of 10 deg$^2$ will greatly reduce the cosmic variance, better prob the source counts of submm sources down to 2$\sim$3 mJy level (i.e., $S/N\approx 4$), recover more obscured star forming population at high redshift, and improve our knowledge of the SFH over cosmic time. 

\subsection{Synergies with other instruments}

The new 850 $\mu$m wide field camera will allow submm surveys of high-z dusty star forming systems in tens to hundreds deg$^{2}$ of sky area with a much improved sensitivity compared to the current SCUBA2 surveys. The observations will serve as the sub-mm counterparts of the large-scale optical, infrared, and radio surveys with future telescopes (such as the LSST, EUCLID, and SKA et c.). This will reveal the obscured star forming over cosmic time that is not detectable in optical/IR. For galaxies at z$\geq$5, the measurement at 850 $\mu$m samples the dust SED close to the peak (see Figure 1). Thus, the combination of 850$\mu$m data with surveys using IRAM/NIKA2 or LMT/TolTEC at longer wavelengths and constraints/upper limits from Herschel/SPIRE at shorter wavelengths will allow the selection of large sample of candidates at very high-z. With the 10 times faster mapping speed using the new 850$\mu$m camera, this new 850~$\mu$m camera will enlarge the current sample of z$>$5 dusty starburst galaxies by nearly two orders of magnitude. And finally with ALMA or NOEMA, we will be able to accurately locate/resolve the SCUBA-2 detections (e.g., \citep{d19}), and spectroscopicly determine the redshift (e.g., \citep{w12}). These will provide a large sample of young galaxies that are in the obscured starburst phase in both field and protocluster environments. Further high resolution imaging with ALMA, NOEMA, and the VLA will probe the gas distribution and kinematics into details \cite{ho16,gu18} and allow a full study of the galaxy evolution and structure formation at the earliest epoch. 

%%%%%%
% Uncomment the following two lines to use Bibtex, then comment out the manual bibliography section, below:
%\bibliographystyle{unsrtMaxAuth} 
%\bibliography{references.bib}  

\begin{thebibliography}{1}

\bibitem[\protect\citeauthoryear{An, et al.}{2018}]{a18} An F.~X. et al. 2018, ApJ, 862, 101

\bibitem[\protect\citeauthoryear{Aretxaga, et al.}{2011}]{a11} Aretxaga I. et al., 2011, MNRAS, 415, 3831

\bibitem[\protect\citeauthoryear{Austermann, et al.}{2010}]{a10} Austermann J. E. et al., 2010, MNRAS, 401, 160

\bibitem[\protect\citeauthoryear{Bertoldi, et al.}{2003}]{b03} Bertoldi F., Carilli, C. L., Cox, P., et al. 2003, A\&A, 406, L55

\bibitem[\protect\citeauthoryear{Bryan}{2018}]{b18} Bryan S. 2018, Atacama Large-Aperture Submm/mm Telescope (AtLAST), p.36

\bibitem[\protect\citeauthoryear{Casey, et al.}{2015}]{c15} Casey C.~M., et al., 2015, ApJ, 808, L33

\bibitem[\protect\citeauthoryear{Casey}{2016}]{c16} Casey C.~M., 2016, ApJ, 824, 36

\bibitem[\protect\citeauthoryear{Carilli, et al.}{2007}]{c07} Carilli C. L., Neri, R., Wang, R., et al. 2007, ApJ, 666, L9

\bibitem[\protect\citeauthoryear{Chapin, et al.}{2011}]{c11} Chapin E. L. et al., 2011, MNRAS, 411, 505

\bibitem[\protect\citeauthoryear{Chiang, et al.}{2017}]{c17} Chiang Y.-K., Overzier R.~A., Gebhardt K., Henriques B., 2017, ApJ, 844, L23

\bibitem[\protect\citeauthoryear{Clements, et al.}{2014}]{c14} Clements D.~L., et al., 2014, MNRAS, 439, 1193

%\bibitem[\protect\citeauthoryear{Clements, et al.}{2016}]{cl16} Clements D.~L., et al., 2016, MNRAS, 461, 1719

\bibitem[\protect\citeauthoryear{Coppin, et al.}{2006}]{c06} Coppin K. et al., 2006, MNRAS, 372, 1621

\bibitem[\protect\citeauthoryear{Cowie, et al.}{2017}]{cowie17} Cowie, L. L., et al. 2017, ApJ, 837, 139

\bibitem[\protect\citeauthoryear{Daddi, et al.}{2005}]{d05} Daddi E. et al. 2005, ApJ, 631, L13

\bibitem[\protect\citeauthoryear{Dannerbauer, et al.}{2014}]{d14} Dannerbauer H., et al., 2014, A\&A, 570, A55

\bibitem[\protect\citeauthoryear{Decarli, et al.}{2017}]{d17} Decarli R., Walter, F., Ba\~{n}ados, E. et al. 2017, Nature, 545, 457

\bibitem[\protect\citeauthoryear{Decarli, et al.}{2018}]{d18} Decarli R., Walter, F., Venemans, B. P. et al. 2018, ApJ, 854, 97

\bibitem[\protect\citeauthoryear{D\'{e}sert, et al.}{2016}]{d16} D\'{e}sert F.-X. et al., Proceedings of the annual meeting of the French Society of Astronomy \& Astrophysics Lyon, June 14-17, 2016, C. Reyl\'{e} et al. (eds.) SF2A, pp.439 - 442, 2016

\bibitem[\protect\citeauthoryear{Dudzevi\v{c}i\={u}t\.{e}, et al.}{2019}]{d19} Dudzevi\v{c}i\={u}t\.{e}, U. et al. 2019, arXiv:1910.07524

\bibitem[\protect\citeauthoryear{Eales, et al.}{2010}]{e10} Eales S. et al., 2010, PASP, 122, 499

\bibitem[\protect\citeauthoryear{Flores-Cacho, et al.}{2016}]{f16} Flores-Cacho I., et al., 2016, A\&A, 585, A54

\bibitem[\protect\citeauthoryear{Geach, et al.}{2017}]{g17} Geach J. E. et al. 2017, MNRAS, 465, 1789

\bibitem[\protect\citeauthoryear{Greenslade, et al.}{2018}]{g18} Greenslade J., et al., 2018, MNRAS, 476, 3336

\bibitem[\protect\citeauthoryear{Gruppioni, et al.}{2013}]{g13} Gruppioni C. et al. 2013, MNRAS, 432, 23

\bibitem[\protect\citeauthoryear{Gullberg, et al.}{2018}]{gu18} Gullberg B. et al. 2018, ApJ, 859, 12

\bibitem[\protect\citeauthoryear{Harikane, et al.}{2019}]{h19} Harikane Y., et al., 2019, ApJ, in press, arXiv:1902.09555

\bibitem[\protect\citeauthoryear{Ho, et al.}{2008}]{h08} Ho L.~C., Darling, J.\& Greene, J. E. 2008, ApJ, 681, 128

\bibitem[\protect\citeauthoryear{Hodge, et al.}{2016}]{ho16} Hodge J. A. et al. 2016, ApJ, 833, 103

\bibitem[\protect\citeauthoryear{Hsu, et al.}{2016}]{h16} Hsu L.-Y., et al. 2016, ApJ, 829, 25

\bibitem[\protect\citeauthoryear{Hughes, et al.}{1998}]{h98} Hughes D. H. et al., 1998, Natur., 394, 241

\bibitem[\protect\citeauthoryear{Leipski, et al.}{2013}]{l13} Leipski C. et al. 2013, ApJ, 772, 103

\bibitem[\protect\citeauthoryear{Livermore, et al.}{2017}]{l17} Livermore, R. C. et al. 2017, ApJ, 835, 113

\bibitem[\protect\citeauthoryear{Madau \& Dickinson}{2014}]{m14} Madau P., \& Dickinson M. 2014, ARA\&A, 52, 415

\bibitem[\protect\citeauthoryear{Magnelli, et al.}{2019}]{m19} Magnelli B. et al., 2019, ApJ, 877, 45

\bibitem[\protect\citeauthoryear{Maiolino, et al.}{2005}]{m05} Maiolino R., Cox, P., Caselli, P., et al. 2005, A\&A, 440, L51

\bibitem[\protect\citeauthoryear{Mocanu, et al.}{2013}]{m13} Mocanu L. M. et al. 2013, ApJ, 779, 61

\bibitem[\protect\citeauthoryear{Miller, et al.}{2018}]{m18} Miller T.~B., et al., 2018, Natur, 556, 469

\bibitem[\protect\citeauthoryear{Muldrew, Hatch \& Cooke}{2015}]{m15} Muldrew S.~I., Hatch N.~A., Cooke E.~A., 2015, MNRAS, 452, 2528

\bibitem[\protect\citeauthoryear{Oteo, et al.}{2018}]{o18} Oteo I., et al., 2018, ApJ, 856, 72

\bibitem[\protect\citeauthoryear{Oliver, et al.}{2012}]{o12} Olive S. J. et al. 2012, MNRAS, 424, 1614

\bibitem[\protect\citeauthoryear{Omont, et al.}{2001}]{o01} Omont A. et al. 2001, A\&A, 374, 371

\bibitem[\protect\citeauthoryear{Omont, et al.}{2003}]{o03} Omont A. et al. 2003, A\&A, 398, 857

\bibitem[\protect\citeauthoryear{P\^{a}ris, et al.}{2018}]{p18} P\^{a}ris I. et al. 2018, A\&A, 613, A51

\bibitem[\protect\citeauthoryear{Planck Collaboration, et al.}{2015}]{p16} Planck Collaboration, et al., 2015, A\&A, 582, A30

\bibitem[\protect\citeauthoryear{Perera, et al.}{2008}]{p08} Perera T. A. et al., 2008, MNRAS, 391, 1227

\bibitem[\protect\citeauthoryear{Priddey, et al.}{2003}]{p03} Priddey R. S. et al. 2003, MNRAS, 339, 1183

\bibitem[\protect\citeauthoryear{Polletta, et al.}{2007}]{p07} Polletta M. et al. 2007, ApJ, 663, 81

\bibitem[\protect\citeauthoryear{Pope\& Chary}{2010}]{p10} Pope A. \& Chary, R. 2010, ApJ, 715, L171

\bibitem[\protect\citeauthoryear{Riechers, et al.}{2013}]{r13} Riechers D. A., Bradford, C. M., Clements, D. L. et al. 2013, Natur. 496, 329

\bibitem[\protect\citeauthoryear{Riechers, et al.}{2017}]{r17} Riechers Dominik A. et al. 2017, ApJ, 850, 1

\bibitem[\protect\citeauthoryear{Roseboom, et al.}{2012}]{r12} Roseboom I. et al., 2012, MNRAS, 419, 2758

\bibitem[\protect\citeauthoryear{Rowan-Robinson, et al.}{2016}]{r16} Rowan-Robinson M. et al. 2016, MNRAS, 461, 1100

\bibitem[\protect\citeauthoryear{Scott, et al.}{2010}]{s10}  Scott K. S. et al., 2010, ApJS, 191, 212

\bibitem[\protect\citeauthoryear{Scoville, et al.}{2014}]{s14} Scoville N. et al. 2014, ApJ, 783, 84

\bibitem[\protect\citeauthoryear{Shangguan, et al.}{2018}]{s18} Shangguan J. et al. 2018, ApJ, 854, 158

\bibitem[\protect\citeauthoryear{Simpson, et al.}{2019}]{s19} Simpson J. M. et al. 2019, ApJ, 880. 43

\bibitem[\protect\citeauthoryear{Smith, Lucey \& Carter}{2012}]{s12} Smith R.~J., Lucey J.~R., Carter D., 2012, MNRAS, 426, 2994

\bibitem[\protect\citeauthoryear{Strandet, et al.}{2017}]{s17} Strandet M. L. et al. 2017, ApJ, 842, L15

\bibitem[\protect\citeauthoryear{Vallini, et al.}{2013}]{va13} Vallini L. et al. 2013, MNRAS, 433, 1567

\bibitem[\protect\citeauthoryear{Venemans, et al.}{2016}]{v16} Venemans B. P., Walter, F., Zschaechner, L. et al. 2016, ApJ, 816, 37

\bibitem[\protect\citeauthoryear{Vieira, et al.}{2013}]{vi13} Vieira J. D. et al. 2013, Nature, 495, 344

\bibitem[\protect\citeauthoryear{Viero, et al.}{2014}]{v14} Viero M. P. et al. 2014, ApJS, 210, 22

\bibitem[\protect\citeauthoryear{Volonteri}{2010}]{v10} Volonteri M. 2010, A\&ARv, 18, 279

\bibitem[\protect\citeauthoryear{Walter, et al.}{2012}]{w12} Walter F. et al. 2012, Natur., 486, 233

\bibitem[\protect\citeauthoryear{Wang, et al.}{2007}]{w07} Wang R., Carilli, C., Beelen, A., et al. 2007, AJ, 134, 617

\bibitem[\protect\citeauthoryear{Wang, et al.}{2013}]{w13} Wang R., Wagg, J., Carilli, C. et al. 2013, ApJ, 773, 44

\bibitem[\protect\citeauthoryear{Wang, et al.}{2017}]{w17} Wang W.-H., et al. 2017, ApJ, 850, 37

\bibitem[\protect\citeauthoryear{Wei\ss, et al.}{2009}]{w09} Wei\ss \ A. et al., 2009, ApJ, 707, 1201

\bibitem[\protect\citeauthoryear{Yuan, et al.}{2014}]{y14} Yuan T., et al., 2014, ApJ, 795, L20

\end{thebibliography}
%%%%%%

%%%%%%
% To construct your bibliography manually:
\bibliographystyle{unsrt}

%%%%%%

\end{document}